\documentstyle[11pt,paspconf,epsf]{article}

\begin{document}

\title{A Complete Sample of Millijansky Radio Sources}

\author{I. Waddington}

\affil{Department of Physics \& Astronomy, Arizona State University, 
Tempe, AZ 85287--1504}

\begin{abstract}

The results of an optical and infrared investigation of a radio sample
drawn from the 1.4~GHz Leiden-Berkeley Deep Survey are presented.
This is believed to be the most comprehensive sample of radio sources
at millijansky flux limits that is currently available.  Optical
counterparts have been identified for all but four sources in the two
Hercules fields, and 80\% of them are identified in the near-infrared.
Redshifts have been obtained for 49 of the identified sources, and
photometric redshifts were computed from the $gri$K data for the
remaining 20.

The general properties of the sample are summarized.  The use of this
sample in measuring the 1.4~GHz radio luminosity function is
discussed.  Finally, {\it HST}/NICMOS images of two old, red galaxies
at $z\simeq 1.5$ are presented which show that both galaxies are
dominated by an $r^{1/4}$ profile with a scale-length of 5~kpc.

\end{abstract}

\keywords{galaxies: active --- galaxies: distances and redshifts --- 
galaxies: evolution --- infrared: galaxies --- radio continuum:
galaxies}

\section{The sample}

The Leiden-Berkeley Deep Survey (`LBDS') consists of nine high
latitude fields in the selected areas SA28, SA57, SA68 and an area in
Hercules.  They were surveyed with the Westerbork Synthesis Radio
Telescope at 21~cm (1.412~GHz), reaching a 5-$\sigma$ limiting flux
density of 1~mJy (Windhorst et~al.\ 1984a).  Multi-color prime focus
photographic plates of the fields were used to find optical
counterparts to the radio sources.  Identifications were found for
53\% of the sources in the full survey, whilst for the Hercules fields
47 out of 73 sources were identified (Windhorst et~al.\ 1984b; Kron
et~al.\ 1985).

The Hercules fields were subsequently observed on the 200~inch Hale
Telescope at Palomar Observatory between 1984 and 1988.  Multiple
observations were made through Gunn $g$, $r$ and $i$ filters over six
runs.  After processing and stacking of the multiple-epoch images,
optical counterparts for 22 of the sources were found, leaving only
four sources unidentified to $r\simeq 26$.  Near-infrared observations
have been made of the entire subsample at K, yielding 60/73 detections
down to ${\rm K}\simeq 19$--21.  Half of the sources have been
observed in H and approximately one-third in J.  Observations of the
brighter sources were made by Thuan et~al.~(1984) and by Neugebauer
et~al.\ and Katgert et~al.\ (priv.\ comm.).  K-band observations of
the sample were completed by the present author and collaborators at
the UK Infrared Telescope.

Prior to the start of the current work, only 16 of the 73 sources in
the LBDS Hercules fields had redshifts published in the literature.
Another 16 sources had unpublished redshifts.  The author and
collaborators have successfully observed a further 17 sources during
the past few years, using both the 4.2~m William Herschel Telescope
(Waddington 1999) and the 10~m Keck Telescope (Dunlop et~al.\ 1996;
Spinrad et~al.\ 1997; Dey 1997).  This brings the total number of
spectroscopic redshifts to 49 out of 73 sources (67\%).  Photometric
redshifts were calculated for the remaining one-third of the sample,
using the spectral population synthesis models of Jimenez
et~al.~(1998).

\section{The 1.4~GHz radio luminosity function}

Dunlop \& Peacock~(1990) used a sample of radio sources brighter than
0.1~Jy at 2.7~GHz to investigate the evolution of the radio luminosity
function (RLF).  They concluded that the comoving density of both
flat- and steep-spectrum sources suffers a cut-off at redshifts
$z\simeq 2$--4.  This conclusion was drawn from the behavior of both
free-form and simple parametric models (PLE/LDE), and the
model-independent, banded $V/V_{\rm max}$ test.  However, the results
were crucially dependent upon the accuracy of their redshift estimates
in the Parkes Selected Regions (PSR).

\begin{figure}
\plottwo{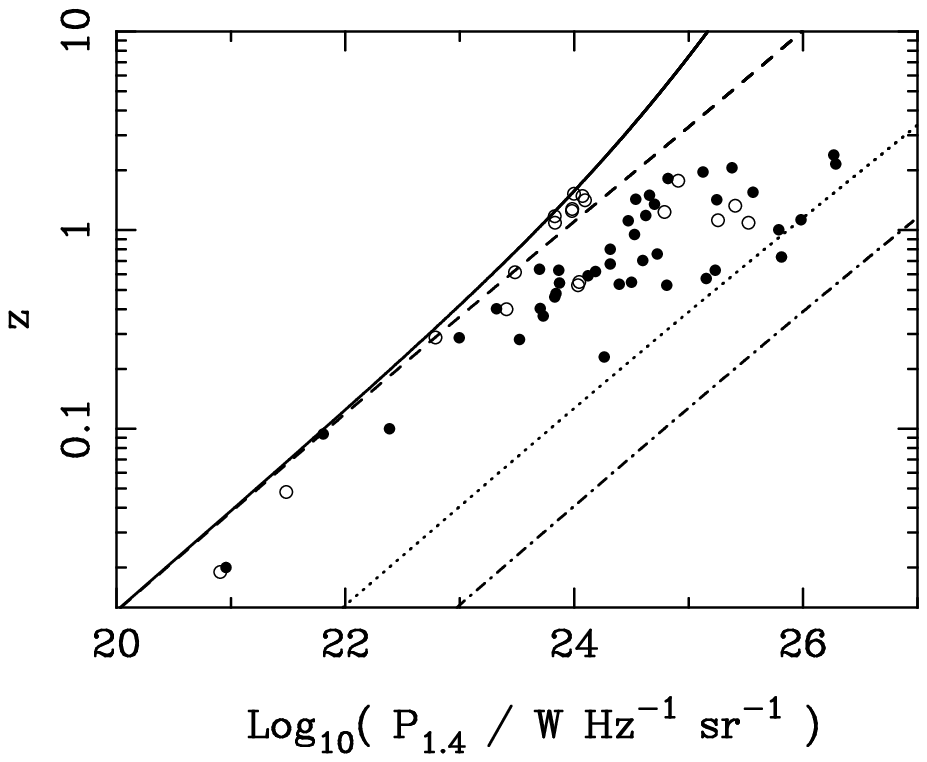}{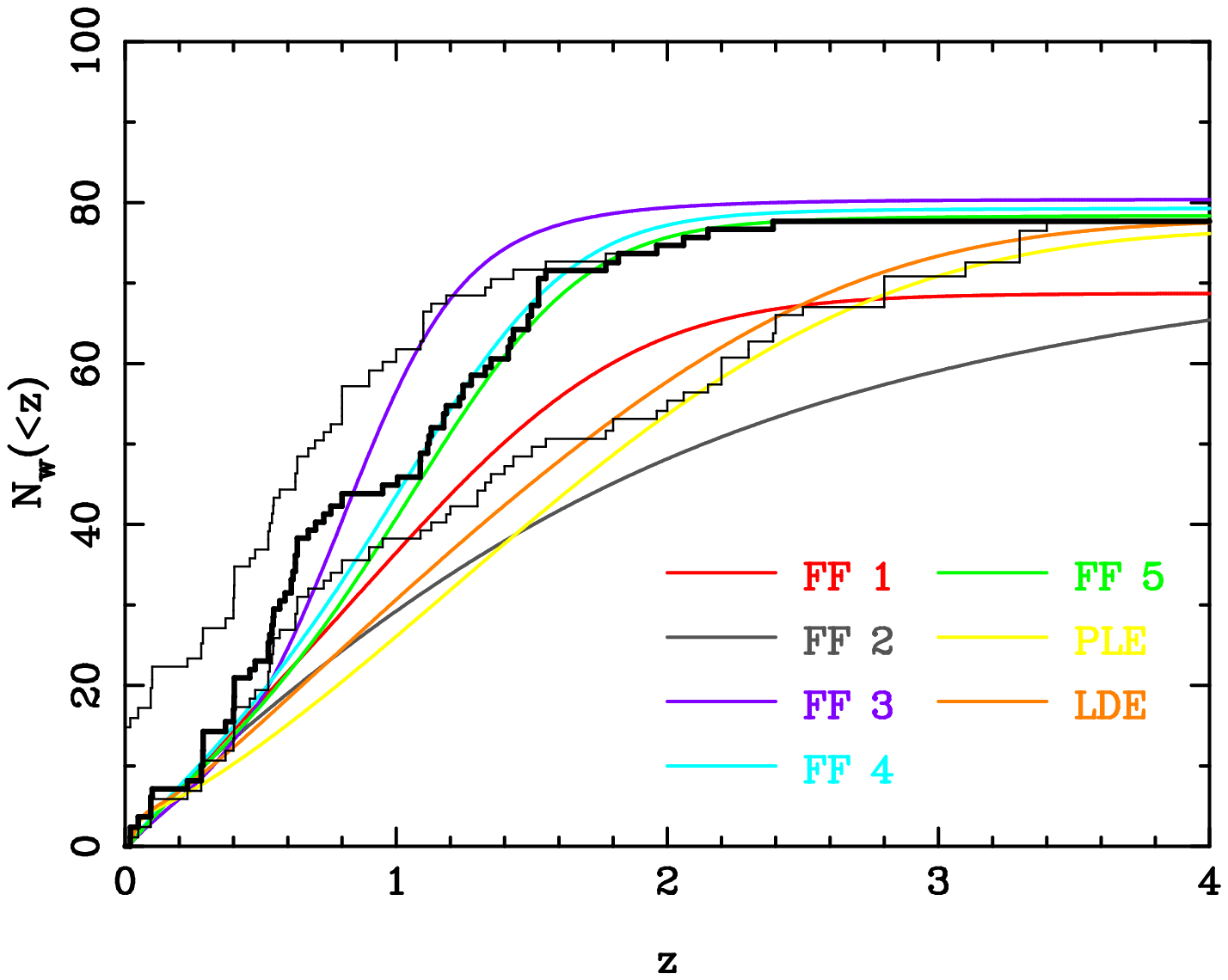}
\caption{(a) [left] The luminosity--redshift plane for sources with
$S_{1.4}\ge 2$~mJy in the LBDS Hercules fields.  The flux density
limits for the survey are shown for flat-spectrum (solid line, open
circles) and steep-spectrum (dashed line, solid circles) sources,
together with the limits for the PSR (dotted line) and 3C (dot-dash
line) surveys. (b) [right] The cumulative redshift distribution for
the $S_{1.4}\ge 2$~mJy sample.  The bold histogram corresponds to the
best-fitting photometric redshift distribution, the light histograms
represent the 99\% confidence limits on the photometric redshifts.
The gray lines are the model luminosity functions of Dunlop \&
Peacock~(1990), assuming H$_0=50$~km~s$^{-1}$~Mpc$^{-1}$,
$\Omega_0=1$, $\Lambda=0$.\label{iwfig1}}
\end{figure}

Figure~\ref{iwfig1}a compares the flux limits of the LBDS with those of
the PSR and 3C surveys.  It is seen how the LBDS can be used to: (i)
probe the faint end of the RLF out to much greater redshifts than the
brighter surveys; and (ii) detect powerful radio galaxies out to very
high redshifts ($\ga 10$).  Thus we are able to use this millijansky
sample to test the reality of the redshift cut-off.  In
figure~\ref{iwfig1}b the cumulative redshift distribution of the LBDS
Hercules sample (only those sources with $S_{1.4}\ge 2$~mJy) is
compared with the predictions of Dunlop \& Peacock~(1990).  It is seen
that two of the free-form models (FF-4 and FF-5) provide a reasonable
fit to the data over all redshifts, suggesting that the high-$z$
decline in the RLF is real.  The ``bump'' in the best-fit histogram at
$0.4\la z \la 1$ is due to two spikes in the redshift distribution,
that may be the result of possible large-scale structures (sheets)
along the line of sight.

However, a more detailed investigation of the RLF reveals that the
{\it luminosity\/} dependence of the data does not agree with the
models.  The observed RLF shows some indication that it turns over at
$z\simeq 0.5$--1.5, and that the redshift of this cut-off is a
function of the radio luminosity.  The small number of sources makes
it difficult to separate the redshift and luminosity dependence of the
RLF sufficiently to be certain of this trend, but work is ongoing to
improve the modeling of the data.

\section{Surface brightness profiles of two old galaxies}

Two of the most interesting individual sources in the sample are
53W069 and 53W091.  Keck spectra of these galaxies revealed that their
restframe ultraviolet light was dominated by old stellar populations
at ages of 4.5~Gyr at $z=1.432$ and 3.5~Gyr at $z=1.552$ respectively
(Dey 1997; Spinrad et~al.\ 1997).  Although there continues to be some
debate over the accuracy of these ages, there can be little dispute
that they are two of the oldest galaxies observed at that redshift.

\begin{figure}
\plottwo{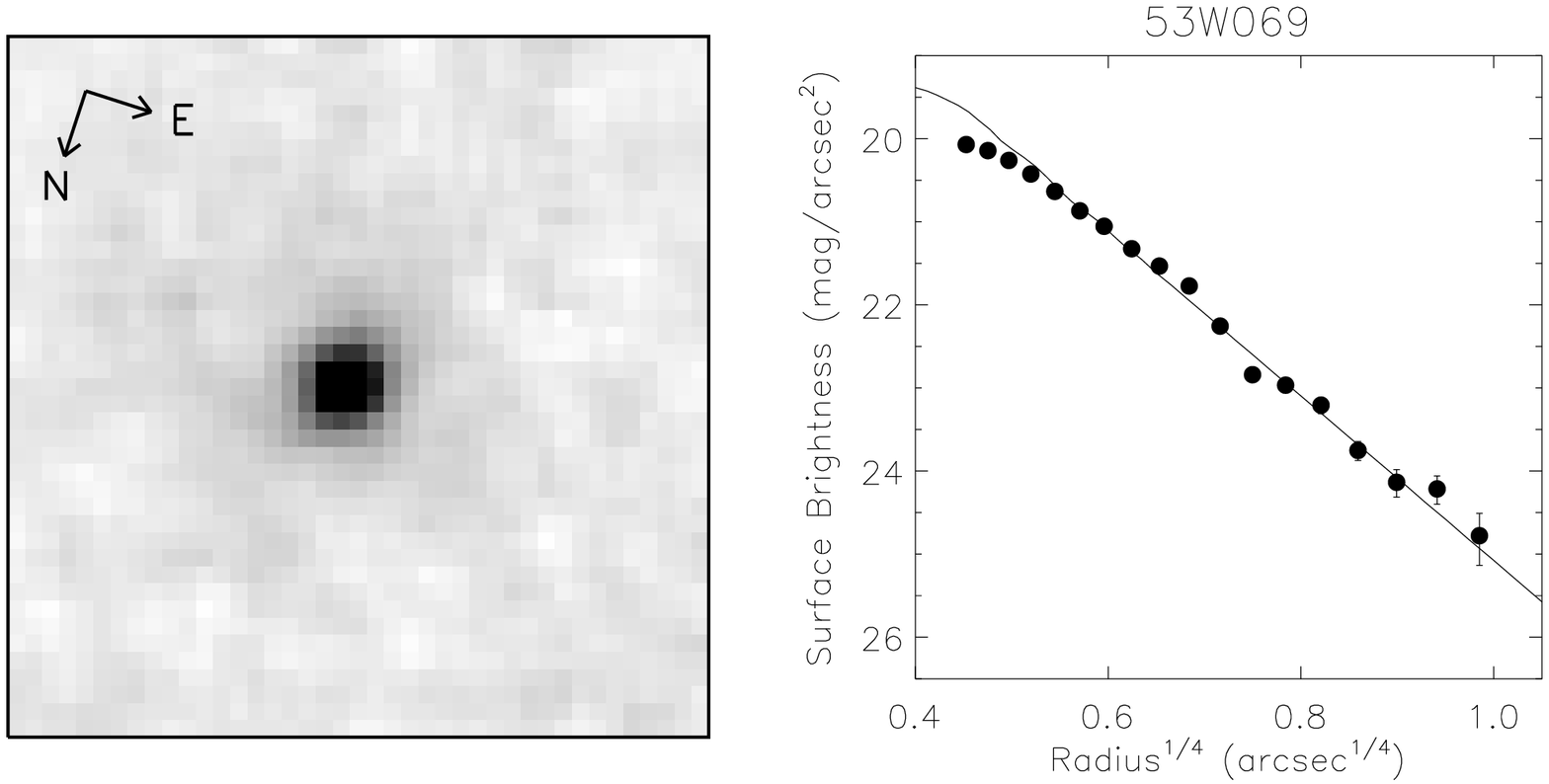}{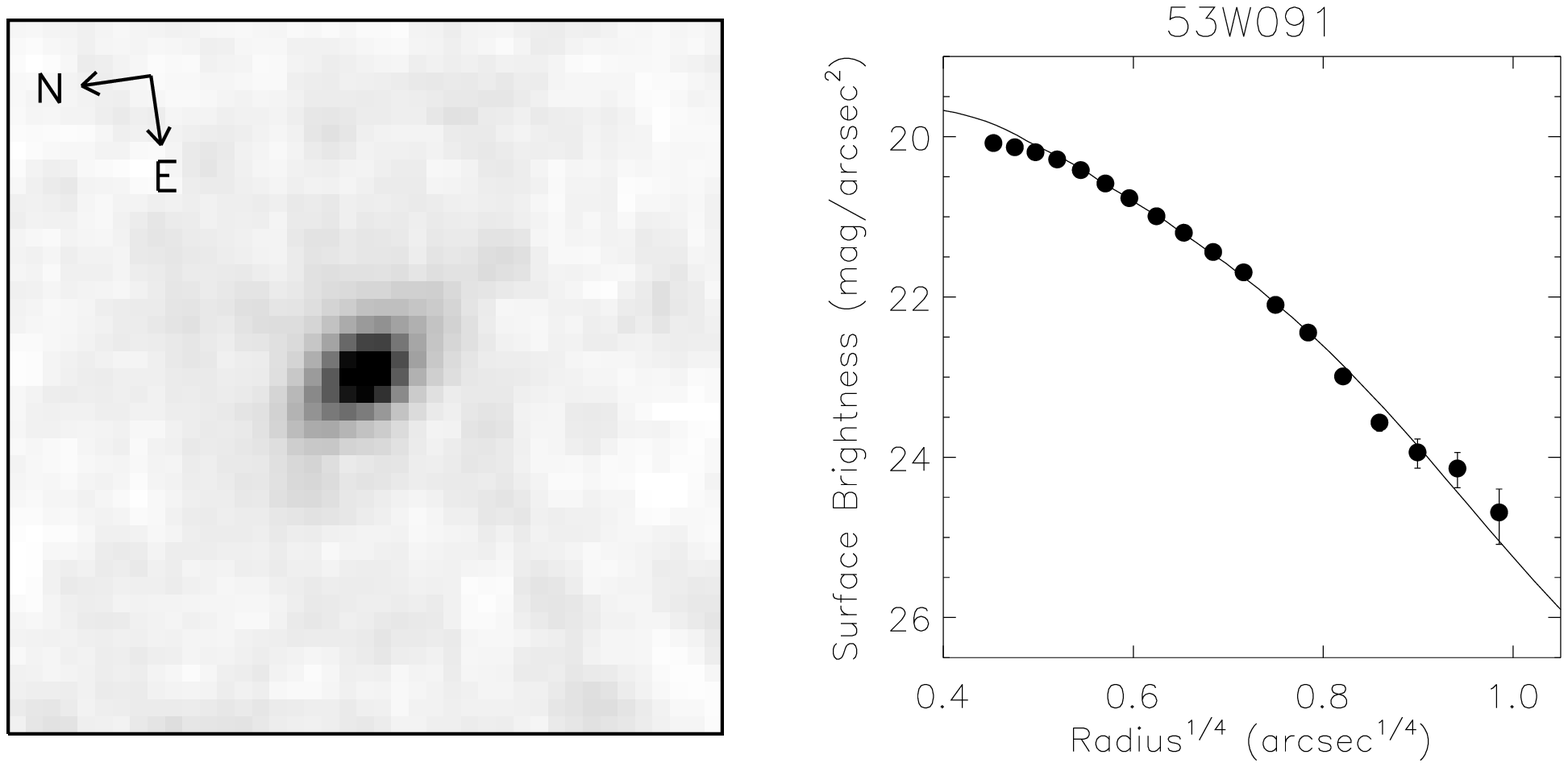}
\caption{{\it HST}/NICMOS F110W images and surface brightness profiles
for 53W069 at $z=1.432$ [left] and 53W091 at $z=1.552$ [right].  The
images are 3\arcsec\ on a side.  Solid lines are the best-fitting
model profiles convolved with the NICMOS PSF.  AB magnitudes are
used.\label{iwnicmos}}
\end{figure}

An important corollary to investigate was whether these galaxies were
also {\it dynamically\/} old objects.  Peacock et al.\ (1998) showed
that in order for the number density of these objects to be consistent
with the evolving power spectrum of primordial density fluctuations,
then the galaxies must be 3--4~Gyr old at $z\simeq 1.5$, having
collapsed at $z\simeq 6$--8.  This result is independent of the ages
of the sources derived from their Keck spectra.  Further clues to the
dynamical history of the two galaxies can be gained by looking at
their morphologies.  53W069 and 53W091 were observed using WFPC2 and
NICMOS on the {\it Hubble Space Telescope\/} in Cycle 7 (Waddington et
al.\ 1999).  Data were collected in F814W and F110W: these filters
bridge the 4000~\AA\ break and are thus most sensitive to the young
and old stars in a galaxy respectively.  Figure~\ref{iwnicmos} shows the
F110W images and surface brightess profiles for the two objects.

53W069 has a regular de Vaucouleurs $r^{1/4}$ profile, with an
effective radius of 0\farcs5 or 4~kpc
(H$_0=65$~km~s$^{-1}$~Mpc$^{-1}$, $\Omega_0=0.2$, $\Lambda=0$).
53W091 is similarly dominated by an $r^{1/4}$ profile of effective
radius 0\farcs5 (4~kpc), however there is an additional component
required to fit this galaxy.  The form of this extra emission has not
been unambiguously identified, but initial results suggest that it is
consistent with an exponential profile of 0\farcs2 (2~kpc)
scale-length, contributing $\sim 40$\% of the F110W flux within a
1\farcs5 diameter aperture.

The spectrum of 53W091 can be modeled by adding a young stellar
population to the spectrum of 53W069.  Similarly the surface
brightness profile of 53W091 can be modeled by adding an exponential
component to the profile of 53W069.  A possible interpretation is that
53W091 has a star-forming disk-like structure surrounding an otherwise
passively evolving elliptical galaxy, such as 53W069.

\acknowledgments I thank my collaborators on this project,
particularly Rogier Windhorst, James Dunlop and John Peacock.  This
work was supported by a research studentship from the UK PPARC; and
NASA grant GO-7280.0*.96A from STScI under NASA contract NAS5-26555.

\end{document}